# Homogeneity of lithium distribution in cylinder-type Li-ion batteries

A. Senyshyn[1,*], M.J. Mühlbauer[2], O. Dolotko[1], M. Hofmann[1], H. Ehrenberg[2,3]

[1]*Heinz Maier-Leibnitz Zentrum (MLZ), Technische Universität München, Lichtenbergstr. 1, 85748 Garching, Germany*
[2]*Institute for Applied Materials (IAM), Karlsruhe Institute of Technology (KIT), Hermann-von-Helmholtz-Platz 1, D-76344 Eggenstein-Leopoldshafen, Germany*
[3]*Helmholtz-Institute Ulm for Electrochemical Energy Storage (HIU), P.O. Box 3640, D-76021 Karlsruhe, Germany*

*Spatially-resolved neutron powder diffraction with gauge volume 2x2x20 $mm^3$ has been applied to probe the lithium concentration in the graphite anode of different Li-ion cells of 18650-type in situ in charged state. Structural studies performed in combination with electrochemical measurements and X-ray computed tomography under real cell operating conditions unambiguously revealed non-homogeneity of lithium distribution in the graphite anode. Deviations from a homogeneous behaviour have been found in both radial and axial directions of 18650-type cells and were discussed in the frame of cell geometry and electrical connection of electrodes, which might play a crucial role in the homogeneity of the lithium distribution in the active materials within each electrode.*

Li-ion batteries are currently dominating the field of electrochemical energy storage especially in portable electronic and electric vehicles applications. Their outstanding position is primarily due to their high power and energy densities, good cycle life and excellent storage characteristics, thus making Li-ion battery technology the favorite on the broad selection of different electrochemical storage media. However there is still space for optimization especially when safety, energy and power, price, weight or temperature is concerned. This caused active research on Li-ion batteries in recent years in line with the rapid development of portable electronics and the general trend substituting fossil fuel by renewable energy. As a consequence cell prices are rapidly falling and an annual increase of ca. 8.5 % for the accessible cell capacity (from ca. 1.2 Ah for first 18650-type cells from Sony in 1991 to about 3.4 Ah in 2013) is achieved.

Strongly application-oriented lithium-ion batteries are now built in different cell designs with the aim to achieve highest energy density. Besides the fact, that a modern Li-ion cell is a sophisticated electrochemical device possessing numerous and intricately coupled degrees of freedom, it may have a complicated shape with different way of electrode layer layout, electrical connection, electrolyte wetting etc. Simple in principle but complicated in practice cell designs may result in spatial inhomogeneity of current, lithium or electrolyte distribution, which are often difficult to quantify, but it will surely affect performance, cycling stability and safety.

A typical lithium-ion battery is an electrically and environmentally isolated stand-alone system, whose thermodynamic or chemical state changes immediately after any interference. This creates numerous challenges for studies of Li-ion batteries and restricts the broad range of experimental and analytic techniques to the use of in-situ non-destructive characterization methods. Hereby risks of possible materials oxidation, electrolyte evaporation, battery state-of-charge (SOC) changes etc. are eliminated. Bearing in mind that a high penetration depth and a sensitivity to light atoms is needed, the applicable methods out of the accessible experimental toolbox become quite limited.

[*] Corresponding author, E-Mail: anatoliy.senyshyn@gmail.com

In such instance neutron scattering is already a well-established experimental technique for the characterization of complex Li-ion batteries. It has numerous advantages[1]: the high penetration depth of thermal neutrons permits non-destructive in situ and in operando studies of different kinds; the capability to localize light elements (e.g. hydrogen, lithium) and to resolve isotopes provides the necessary contrast; the neutron scattering length not dependent on momentum transfer, enabling to achieve accurate structure factors at high 2θ angles etc. Neutron scattering methods undergo an increasing relevance for studies of lithium-ion batteries on different length scales, e.g. neutron imaging[2-4], reflectometry[5], small-angle neutron scattering[6-8], quasielastic neutron scattering[9] and powder diffraction[4, 10-15]. In-situ experiments with neutrons are performed on self-developed/special test cells and commercial Li-ion cells of different designs depending on the research needs. When neutron powder diffraction is considered the adapted coin-type cells[16], pouch-back cells[17], Swagelok-based constructions[18-19], cylinder winded cells[20-22] or commercial design of 18650[4]/prismatic[10] types are often utilized with the main focus on the development of atomic structure of electrode materials upon various electrochemical and environmental states.

Despite the increasing popularity of neutron scattering studies of batteries at their operating conditions (especially neutron powder diffraction where typically large cell cross sections are exposed to the neutron beam) the problem of cell homogeneity and its effect on the data is often not properly accounted in literature. However the fact that characteristic peaks at plateaus on differential capacity plots (cycled under similar conditions) for cells in 2032 standard (coin-type) are narrower than their counterparts from 18650-type cells indirectly points out a possible inhomogeneity issue in larger format Li-ion cells[23]. This can be associated with the development of different state of charges in large Li-ion cells due to the voltage drop along the current collector during charge[24]. To our best knowledge structural studies of cell homogeneity are limited to the work of Cai et al.[25]. They revealed an inhomogeneous deterioration at the edges of a pouch-back cell after active cycling. Recently[26] we reported results of spatially resolved monochromatic neutron powder diffraction applied to commercial Li-ion cells of 18650-type and LCO cathode along the radial axis where the uniformity of Li distribution in the cell volume was concluded along the studied axis. However further experiments performed in other radial directions and various types of cells showed that this not always holds true and requires more systematic characterization.

In the current manuscript we report a further study of the lithium distribution in the graphite anode of different cells of 18650-type by using spatially resolved neutron powder diffraction. Four types of commercial cells based on graphite anodes and different cathode materials were chosen for the experiment and their properties are listed in Table 1 with their applied voltage windows and nominal capacities.

**Table 1. Details of the 18650-type cells studied.** There <x> corresponds to mean lithium content x in the $Li_xC_6$ obtained after averaging over all investigated gauge volumes and $x_P$ defines the plateau with homogeneous lithium distribution (see below).

| Cell Nr | Composition Cathode\|Anode | Capacity, mAh | Voltage window, V | Mean lithium content $x$ in $Li_xC_6$ Mean <x> | Plateau $x_P$ |
|---|---|---|---|---|---|
| Cell 1 | $LiCoO_2$\|C | 2600 | 3.0 -4.2 | 0.85(4) | 0.88(1) |
| Cell 2 | $LiNi(Co, Mn)O_2$\|C | 2200 | 2.75 -4.2 | 0.77(3) | 0.82(1) |
| Cell 3 | $LNi(Co, Al)O_2$\|C | 3400 | 2.5-4.2 | 0.87(5) | 0.93(1) |
| Cell 4 | $LiFePO4$\|C | 1100 | 2.0 -3.6 | 0.68(5) | 0.77(1) |

The layout of cells 1-4 was non-destructively probed using X-ray tomography, where results of the tomography reconstruction of X-ray attenuation for cell 1 are shown in Fig. 1a-b, d-f. Due to poor sensitivity of X-ray based methods to light atoms the lithium distribution in the anode was studied using spatially resolved neutron diffraction. There by translating and rotating the sample with respect to the fixed gauge volume a spatially-resolved scattering response from different sample sections can be obtained (see Methods section and Fig. S1 for details). All investigated cells (Nr. 1-4) were studied in charged state. Scans at the center of the cells were performed in a base cylinder plane (Fig. 1a). The distribution of the studied gauge volumes (red rhombs) is shown in Fig. 1b. A diffraction pattern similar to that displayed in Fig. S1c was collected at each point. The lithium concentration in $Li_xC_6$ was determined using relative intensities of 001 $LiC_6$ and 002 $LiC_{12}$ Bragg reflections. Surfaces of spatial $x$ distribution were constructed by the interpolation of experimental data using biharmonic spline interpolation[27] (see Fig. S2 for details) and are shown for cell 1 in Fig. 1c using false color presentation with $x$ ranging from 0.5 (blue) to 0.95 (red).

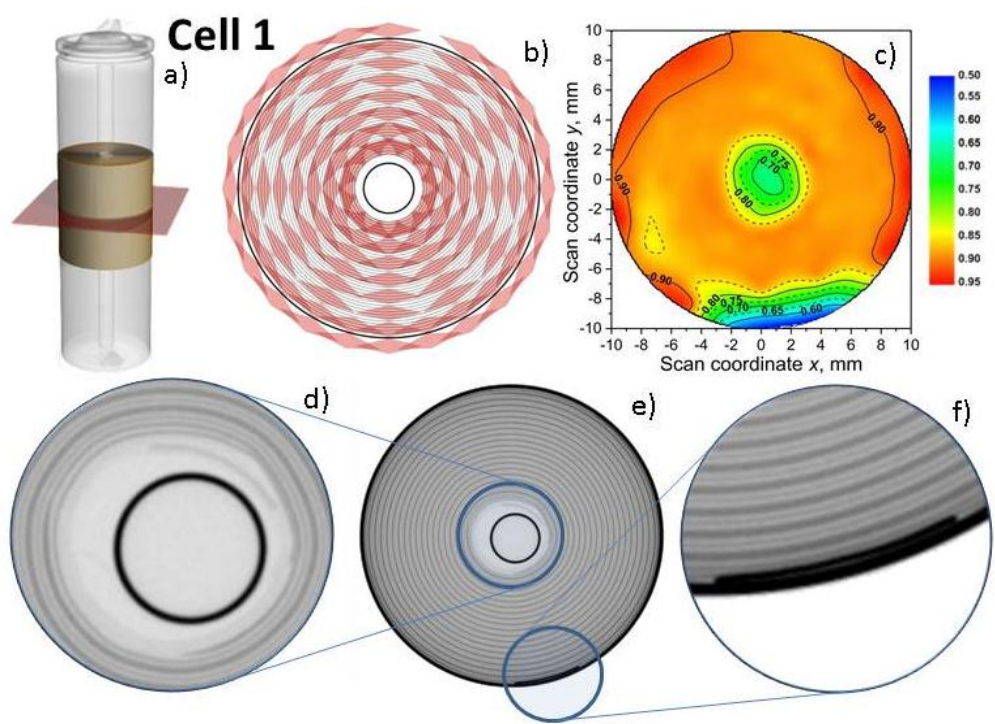

**Figure 1. 3D organization and lithium distribution in cell 1.** (a) Colored part indicating the range of interest for spatially-resolved neutron powder diffraction, red plane defined by the center of incoming and scattered neutron beam; (b) planar distribution of gauge volumes (20 mm in height); (c) planar distribution (2D) of lithium concentration x in $Li_xC_6$ for cell 1 (see Fig. S2 for details). (d-f) Slice of 3D tomography reconstruction and enlarged sections from red plane in a) (grayscale correspond to inverse absorption).

In order to derive the lithium concentrations the analysis of relative peak intensities has a number of advantages compared to a previously reported treating of relative intensities of single peaks[25-26], e.g. the effect of absorption can be neglected in first approximation, lithium inhomogeneities will become apparent in form of deviations from the flat surface and no signal is expected when no material is in the beam (within the studied gauge volume). From Fig. 1 one can see that for cell 1 the 2D map of the lithium distribution shows homogeneous character besides the narrow regions at the center and close-to the surface. For the sake of comparison two lithium concentrations inside the anode will be analyzed, namely $x_P$ given by the plateau with homogeneous lithium distribution and the mean lithium content $<x>$ obtained after averaging over all investigated gauge volumes. For cell 1 a fair agreement between $x_P=0.88(1)$ and $<x>=0.85(4)$ was observed (see Table 1). One has to admit that in areas with reduced lithium concentrations the couple of 001 $LiC_6$ and 002 $LiC_{12}$ reflections were clearly observed (Fig. S2) indicating the presence of both stage I and stage II lithiated carbons in the corresponding gauge volume.

The existence of regions with low lithium concentrations can be explained in the frame of a typical design of cylinder cells adopting double-coated electrodes wound around a center pin. A reconstructed slice form the X-ray tomography corresponding to the diffraction plane is shown in Figs. 1d-f. In case of cell 1 the typical sequence of the electrode layers is "… / Cu/ C| S| $LiCoO_2$/ Al/ $LiCoO_2$| S| C/ Cu/ C| S| $LiCoO_2$/ Al/ …", where S is the separator, $LiCoO_2$/ Al/ $LiCoO_2$ corresponds to the cathode coated on both sides of aluminium foil and C/ Cu/ C is the graphite anode on copper foil, respectively. Due to the necessary connection to current leads the electrode foils are not coated at their ends (Fig. 1d and 1f). The coating of opposite electrodes ($LiCoO_2$/ Al/ $LiCoO_2$ and C/ Cu/ C) usually does not start at the same winding position. This may result in layer sequences like "… S| C/ Cu/ C| S| C/ Cu/ C| S …" or "… S| Al| S| C/ Cu/ C| S …" where there is no cathode in the neighbouring layer and, correspondingly, no lithiated graphite. A similar feature may be observed on the cathode sheet. It has to be noted that no traces of the graphite 002 peak (theoretical position 27.78 deg. 2θ) were observed in the current study. This along with the structural evidence for lithiated carbons can be associated with sufficiently high intralayer lithium diffusion along the negative electrode or a sufficient distribution and high ionic conductivity of the electrolyte. The observed lithium deficiency in cell 1 at regions close to current leads connections, the center pin and the outermost layer of the cell may be explained by a lack of cathode material, which is only partially compensated by the ionic conductivity of the electrolyte and inside the electrode.

Similar experiments were performed on other representatives of 18650-type cells. Cell 2 has been found to possess a similar design to that of cell 1 (as confirmed by X-ray computed tomography) and delivered similar spatial distribution of lithium in the graphite anode (Figs. 2a). A lower grade of graphite lithiation was observed in cell 2, which can be attributed to different cell balancing having a higher excess of graphite. This is also reflected in a lower cell capacity. A flat concentration plateau with $x_P=0.82(1)$ was observed in a (narrow) range of the cell with radii between 3 mm and 7 mm around the centre pin (Fig. S3). At the outer part of the cell a lower lithium concentration in the anode can be identified (Fig. S6a), which causes a higher discrepancy between $x_P$ and $<x>=0.77(3)$ and accordingly indicates a less homogeneous lithium distribution in cell 2 compared to cell 1

Cell 3 was characterized by the highest capacity in the series of studied cells. From Figs. 2b, S4 and S6c it can be seen that the plateau value $x_p=0.93(1)$ is the highest in the series, which correlates well with the cell capacity. Cell 3 also possesses a remarkable difference in design (Fig. 2b) – its positive current lead (connecting to the cathode) is located in the middle of the electrode and therefore positioned between the layers of the rolled stack of electrodes. This geometrical detail has a pronounced effect on the homogeneity of the lithium distribution in the graphite anode. At the location of current leads the lithiation level has been found definitely lower: 0.58(1) at the negative (anode) current lead (outer part) and ca. 0.77(1) at the positive (cathode) current lead located in between the rolled electrodes at a radius of approximately 6 mm from the center of cell 3[†]. This results in a high mean lithium concentration in the anode of cell 3 with <x>=0.87(5) and in a stronger inhomogeneity compared to cells 1 and 2.

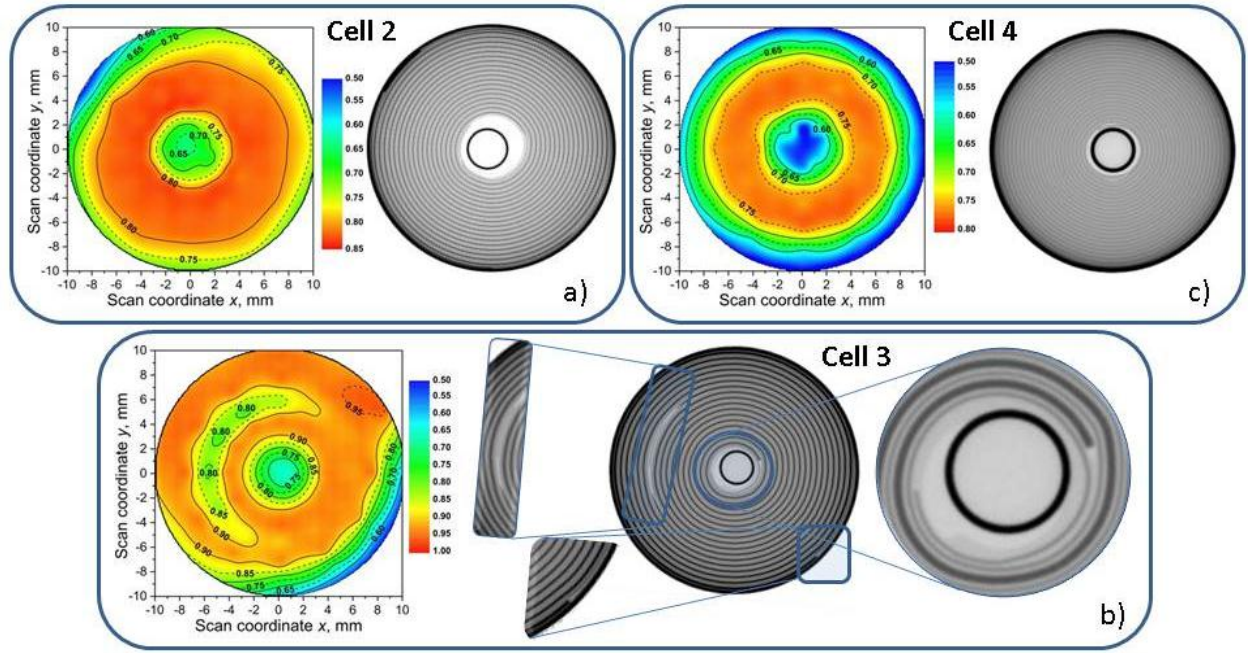

**Figure 2. The lithium concentration $x$ in $Li_xC_6$ and reconstructed X-ray attenuation maps for cells 2-4.** Cell 2 (a, Fig. S2), cell 3 (b, Fig. S3) and cell 4 (c, Fig. S4) at the cell center. Experimental setup and the gauge volume distribution were chosen similar to Fig. 1 a, b.

Even more pronounced deviations from a "flat" lithium distribution were noticed for cell 4 (Figs. 2c, S5, S6d) possessing the lowest cell capacity among all investigated cells. A flat "plateau" with relatively low $x_P$=0.77(1) occurs in a narrow range of radial distances from 4 mm to 7 mm from the centre pin, besides which the lithium concentration rapidly decreases to x ~ 0.50 – 0.55[‡]. A large drop of the lithiation level in the inner and outer regions of cell 4 causes a relatively low mean lithiation state of graphite <x>=0.68(5). The cell is therefore characterized by a high standard deviation of the lithiation level x, corresponding to a strong inhomogeneity of lithiation.

Comparing the <x> and $x_p$ values obtained for cells 1-4 and taking their relative difference as a measure of lithium homogeneity one may notice that the lithium homogeneity is not correlated to the cell capacity and decreases in the series 1 > 2 ≈ 3 > 4, i.e. the cell 1 is characterized by the best value of lithium homogeneity, the overall homogeneity of cells 2 and 3 is nearly similar and the lithium distribution in cell 4 shows the largest deviations from the plateau.

---

[†] The lithium concentration in the anode has been found slightly perturbed at the opposite side (red plot in Fig. S6c) of the cell at ca. 6 mm radial distance from the center of cell 3 (marked by a blue arrow in Fig. S6c). This can be attributed to an inhomogeneous rolling due to a thickness misfit of the current lead and the electrode layers.

[‡] In cells 1 and 2 x~0.55 was observed locally in regions close to the current leads, whereas in cell 4 a radial distribution can be concluded.

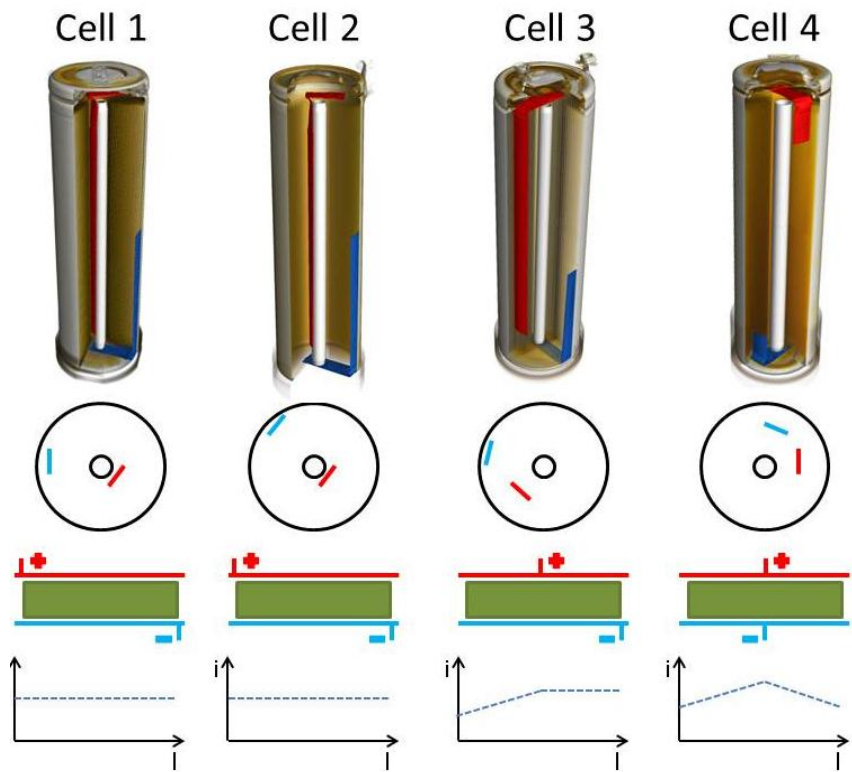

**Figure 3. Electrical connection schemes and current distributions for cells 1-4.** Top: 3D tomography of cells 1-4 illustrating the different configurations of the current leads inside the investigated cells. Positive and negative current leads are shown by red and blue. Middle: The corresponding model of the electrode configuration for the unrolled electrodes. Bottom: Expected current distribution versus position l along the length of the electrodes.

Differences in the lithium concentration profile observed in all four cells of 18650-type can be attributed to the scheme of the electrode connections leading to a non-uniform current distributions. Similar effects were already discussed in literature formerly[23-24]. Depending on the position of the positive and negative tab and the contribution of the current collectors one may expect different resistive paths for the current within a cell. Given that the current collectors possess a similar resistivity per length there is a linear increase of the total resistance of a current path within the current collector with larger distance from the tab position. Three different schemes were found for the investigated cells:

**Scheme 1** (realized in cells 1 and 2): The positive tab is positioned at the center of the cell corresponding to the end of the cathode current collector, while the negative tab is placed at that end of the anode current collector ending at the outer cell section (Fig. 3). For an unrolled electrode stack one would find the positive tab on one end and the negative tab at the other end. For this configuration the total resistance of any current path from the positive tab to the negative tab is independent of the position l, where the electrode layer is crossed. From this one would expect a homogeneous current distribution/density i versus l. Likewise this configuration should result in a homogeneous lithiation of the graphite during charge and therefore in the charged state of the cell.

**Scheme 2** (cell 3): In contrast to scheme 1 the positive current lead is located at the middle of the cathode layer (Fig. 3). For the part right of the positive tab in the sketch of the unrolled electrodes a configuration similar to scheme 1 is achieved, where the resistance is independent of the current path from the positive to the negative tab. Therefore one expects a constant current distribution for this part of the cell, from the position of the positive tab to the outside of the cell. However, on the left side of the positive tab the resistance is increasing for positions closer to l=0 or in other words towards the center of the cell. Due to an increasing resistance with the length of the current path reaches a maximum at the center of the cell (left end of the electrodes). Thus a decrease of the current density I is expected, when moving from the positive tab to the left end of the electrodes. This scheme suggests some lithium inhomogeneity towards the center of the cell. Actually a reduced lithiation is observed at the center of the cell. But there are larger effects caused by the missing coating of the cathode at the position of the positive current lead.

**Scheme 3** (cell 4): For this scheme both leads are positioned in the middle of the electrode stripes (Fig. 3). On the one hand this configuration seems to be the most appropriate for high power cells, as it offers a minimum overall resistance between the positive and the negative leads. The maximum length of a current path is only reached for positions at the end of the electrodes. On the other hand it causes the largest deviation for the total resistance of current paths at the outer ends of the electrodes (left and right) or close to the positive and negative tab (middle). During cell operation (either charge or discharge) it may result in an inhomogeneous current distribution, with a maximum at the middle of the electrode layers and minima at both ends of the electrodes. Accordingly a certain lithium inhomogeneity can be expected proportional to the current distribution, and, therefore, higher litihiation levels close to the leads and lower ones at the outside and center of the cell. Although one would expect some kind of equilibration during the rest phase after charging the cell, the measured lithiation levels inside the cell still reflect a behaviour matching the modeled current distribution.

In summary, the homogeneous behavior of cell 1 and cell 2 is then consistent with the contact Scheme 1, the inhomogeneities observed in cell 4 are explained by adopting contact Scheme 3 and for cell 3 a behavior resembling a mixture of cell 1 and cell 4 can be realized. This holds to be not completely true: cell 2 shows remarkable "wings" at the outer cell regions, whereas cell 3 is characterized by a flat (plateau-like) lithium distribution where inhomogeneities are primarily caused by current leads. On the other hand the proposed contact model initially predicts homogeneous lithium distribution in vertical cell direction (along the cylinder axis).

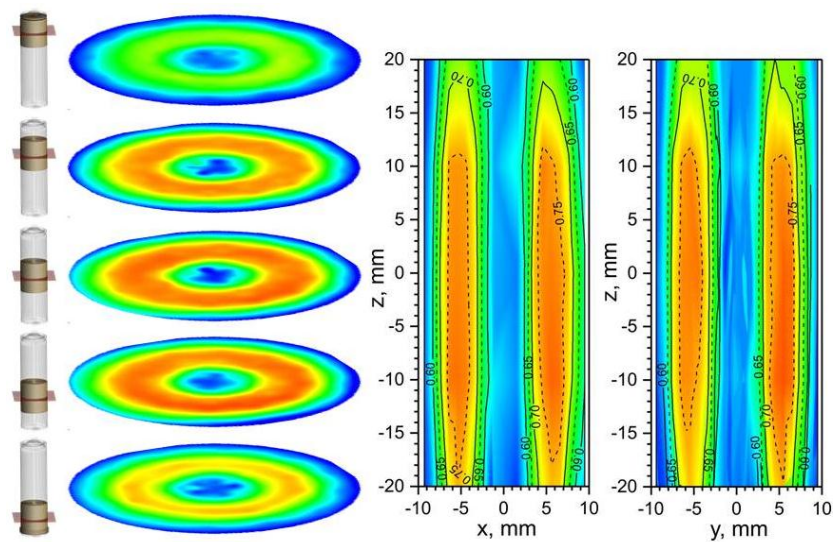

**Figure. 4. Lithium concentration in graphite anode of cell 4 at different heights.** (Left) Acceptance range (range of interest) of spatially-resolved diffraction experiment. The gauge volume distribution was chosen similar to Fig. 2b. (Middle) Lithium concentration x in Li$_x$C$_6$ of cell 4 determined at different heights. (Right) Reconstructed vertical distribution of lithium concentration in cell 4. Slices of experimental data are shown in Fig. S7.

In order to prove this the lithium distribution in cell 4 it was probed at different heights of the sample with respect to its center (-20 mm, -10 mm, 0 mm, +10 mm and +20 mm) using unchanged beam geometry and collimation. A smaller slit (2x10 mm$^2$ instead of 2x20 mm$^2$) for the incoming neutron beam was used in order to increase the spatial resolution in z-direction (along the cylinder axis). The scan at 0 mm is consistent to the dataset shown in Fig. 2c. Results of the z-scans in form of radial lithium distribution at different heights and z-x (z-y) slices are presented in Fig. 4 according to the previously used color scheme.

In general the scans performed at different heights obey a similar behavior to that at 0 mm: a narrow (flat) plateau for radial distances from 4 mm to 7 mm with respect to the centre (pin), beyond which the lithium concentration x in Li$_x$C$_6$ rapidly decreases to x ~ 0.50 – 0.55. However, the observed magnitude of the concentration plateaus has been found clearly depending on the cell height (Fig. S7): plateaus at ca. 0.76(1) have been noticed for scans at cell heights of -10 mm, 0 mm and +10 mm. Slightly lower graphite lithiation with a maximum concentration of ca. x=0.74 was observed for the slice at a cell height of -20 mm, whilst the corresponding scan performed at +20 mm revealed a maximum lithiation of x=0.68 only. A significantly smaller lithium concentration in the graphite anode has been found at the top and bottom parts of cell 4, which might be attributed to a mismatch between electrode (cathode and anode) stripes. This indicates an asymmetry of lithium distribution in the anode occurred in cell 4 at top (+20 mm) and bottom (-20 mm), where lithium concentration in Li$_x$C$_6$ lithiated graphites differ by 0.06. On the first glance it can be attributed to a gravity effects on the electrolyte concentration in the working cell, i.e. its flowing off, but the genuine reason for this phenomenon requires further research.

## Conclusions

The homogeneity of the lithium concentration in the graphite anode (in charged state) of 18650-type cells with different electrode materials and electrode connection schemes was probed using spatially-resolved neutron powder diffraction. The lithium concentration *x* in lithiated graphites Li$_x$C$_6$ was determined from the relative intensities of 002 and 001 reflections of LiC$_{12}$ and LiC$_6$, respectively.

The lithium distribution in the considered 18650-type cells can be characterized by a constant (plateau-like) behaviour where the magnitude (height) of the plateau has been found in good agreement with the cell capacity. In general deviations from the constant plateau-like shape were observed at the regions close to the center pin, the outside of the cells and around current leads, which was associated with the lack of electrode coating for the neighboring cathode sheet. However in some cells deviations from a plateau-like behavior were noticed also in the regions with sufficient electrode overlap (as revealed by X-ray computed tomography), e.g. in cell 4 the constant concentration range was only observed in a very narrow range of radial distances (4 – 7 mm) with respect to the center (pin). Similar observations hold for cells 2 and 3, but have been found less pronounced. Observed inhomogeneities of lithium distribution in the anode were analyzed with respect to the connection of electrode stripes, which may severely influence the lithium distribution. Behaviour of cells 1 and 4 can be fairly explained, whereas behavioral details of cells 2 and 3 as well as observed lithium inhomogeneity at different slices/heights of cell 4 can not be properly accounted. Spatial lithium homogeneity might depend not only on the spatial current distribution, but also on electrode materials with complicated morphology, electrolyte, wetting, state and type of solid-electrolyte interphase, various geometrical factors etc. Information about underlying processes defining lithium distribution is crucial for the manufacturing of safe, robust and high-performance Li-ion cells. It has to be noted that other contributions like current tabs inside the electrode rolls (as shown in cell 3) cause irregularities of the electrode coating and an inhomogeneous winding. In combination with a missing coating of the electrodes at both ends this has an obvious effect on the lithium distribution.

In summary, spatially resolved neutron powder diffraction proves to be a powerful tool for probing the lithium distribution in graphite and has sufficient potential to probe cell homogeneity in various cell designs and states.

## Methods

*Electrochemical characterization.* Prior to diffraction and tomography experiments "fresh" (formed) commercial cells (Nr. 1-4 types) were "re-activated" by charging and discharging in the nominal voltage windows (Table 1). Cell cycling was performed using a CCCV protocol at a constant current of 400 mA; In order to achieve quasi-equilibrium cell conditons the cutoff current in the CV phase of discharge was set to 1 % of 1C. The battery charge/discharge was applied using a VMP3 multichannel potentiostat from Bio-Logic and obtained cell capacities were found in fair agreement with the nominal values listed in Table 1.

*Computed X-ray tomography.* The internal cell geometry with a focus on the current leads was checked by X-ray computed tomography using a v|tome|x s 240 tomography scanner from GE. The direct tube was used at a voltage of 130 kV and a current of 100 µA. The measurement was carried out in cone beam geometry with a sensitive detector area of 200 mm x 200 mm corresponding to 1000 x 1000 pixels. The magnification was adjusted to get an effective pixel size of 40 µm. For every sample two data sets of 1000 projections each were collected, where each projection corresponds to an average of 3 to 4 single exposures of 1000 ms. The reconstruction was performed with the phoenix datos|x software. Two 3D datasets (top and bottom part) were merged for every cell using ImageJ. The final visualization of the data including the separation and colour coding of the different parts was performed in VGStudio MAX from VolumeGraphics.

*Spatially-resolved neutron powder diffraction.* The spatially-resolved neutron powder diffraction study was performed at the engineering diffractometer STRESS-SPEC[28] equipped with translation and rotation stages for sample positioning with respect to the direct and scattered neutron beam. A sketch of the experimental setup is displayed in Fig. S1a. On the first instance a cylindrical cell of the 18650-type can be approximated by two concentric cylinders with an absorbing and scattering medium in between, where the outer cylinder coincides with the steel housing and the inner one is related to the centre pin. Usually the active electrode layers are isolated from the cell housing and the centre pin by several layers of separator. Assuming an isotropic medium, which holds to be fairly correct[26] for gauge volumes larger than 1 mm$^3$, the spatial distribution of structural details of the cell organization can be monitored using neutron powder diffraction on the given length scale.

The shape of the gauge volume was defined by the collimation of the incident and scattered neutron beam and the scattering angle. The incident monochromatic neutron beam (25' horizontal beam divergence, $\lambda$=1.615 Å) was shaped to 2 mm width and 20 mm height using the slit system before the sample. The scattered signal was analyzed by a radial oscillating collimator with acceptance width of 2 mm. The 2D position sensitive neutron detector (300 x 300 mm$^2$ active area, 1056 mm sample-to-detector distance, central scattering angle 26 deg. 2$\theta$) was used for data collection. Diffraction measurements were performed against time, where the exposure time per pattern was in the range of minutes. A typical 2D diffraction dataset showing sections of Debye-Scherrer rings for two reflections is displayed in Fig. S1b. A homogeneous intensity distribution along the diffraction rings in vertical direction was observed thus indicating that there is no prominent/clear contribution of preferred orientation to the diffraction intensities. The 2D diffraction data were corrected for detector nonlinearities, geometrical aberrations and curvature of diffraction rings. After that they are transformed into conventional 1D diffraction patterns (intensity vs. 2$\theta$ angle, Fig. S1c) by data integration.

*Analysis of spatially-resolved neutron powder diffraction data.* For powder diffraction data collection a central detector angle of 26 deg. (2$\theta$) was chosen in order to map the structural evolution of 00*l* reflections from a mixture of stage I (LiC$_6$) and II (LiC$_{12}$) compounds. For these chemical states of the anode the structure remains unchanged and the relative phase concentration is enough to determine the lithium content in the graphite anode and, correspondingly, the local state of charge. Being extremely sensitive to state-of-charge, state-of-health and temperature the stage I/stage II ratio is often used for monitoring Li-ion cell performance[12-13,25,26,29]. Molar quantities were determined from the diffraction data as $M=I_{hkl}[V/F_{hkl}]^2$, where $I_{hkl}$ is an observed intensity of the Bragg reflection with Miller indices *hkl*, $F_{hkl}$ is the corresponding structure factor and $V$ is the cell volume. The couple of 001 LiC$_6$ and 002 LiC$_{12}$ reflections (defining the interplanar distances between the graphite sheets in corresponding lithium intercalated carbons) is well suited for the quantization of relative phase fractions of stage I and II due to their high-intensity and sufficient separation (no reflection overlap). Using the structure factors $F_{001}$= 4.197 (LiC$_6$) and $F_{002}$= 8.022 (LiC$_{12}$) along with the cell volumes $V_{LiC6}$= 59.676(8) Å$^3$ and $V_{LiC12}$= 112.284(9) Å$^3$ the LiC$_6$ to LiC$_{12}$ molar ratio ($n_{LiC6}$:$n_{LiC12}$) for the case shown in Fig. S1c can be estimated as 0.863(9):0.137(9). However, for the sake of comparison it is more comfortable to deal with a single parameter and to switch to a relative (mean) lithium concentration $x$ in the lithiated graphite Li$_x$C$_6$ defined as $x=6/(12-6n_{LiC6}) = 1/(2-n_{LiC6})$. This will yield $x=1$ in Li$_x$C$_6$ for a pure stage I with $n_{LiC6}=1.0$, whilst for a pure stage II $x=0.5$[§]. For the LiC$_6$ to LiC$_{12}$ ratio shown in Fig. S1c $x=0.880(7)$ in Li$_x$C$_6$. The standard uncertainty $\Delta x$ in Li$_x$C$_6$ has been found strongly dependent on the Bragg peak intensities so that $\Delta x$ typically ranges from 0.002 to 0.032 (mediated by neutron attenuation) with $<\Delta x>=0.01$. The angular dependence of neutron transmission may only slightly influence the intensity ratio of neighbouring Bragg reflections not exceeding $<\Delta x>=0.01$ for a linear attenuation coefficient of $\mu=0.221$ mm$^{-1}$.

---

[§] The stability range for stage II is quite broad ($x=0.25-1.0$)[12] thus leading to a poor definition when $x \leq 0.5$. In the current study traces of 001 LiC$_6$ reflection were always present, indicating $x>0.5$ in Li$_x$C$_6$ hereby.

## Author contribution

## Competing financial interests:

The authors declare no competing financial interests.